# Work function and surface stability of tungsten-based thermionic electron emission cathodes

Ryan Jacobs[1], Dane Morgan[1], and John Booske[2]

[1] Department of Materials Science and Engineering, University of Wisconsin-Madison, Madison, WI, USA
[2] Department of Electrical and Computer Engineering, University of Wisconsin-Madison, Madison, WI, USA

**Abstract**

Materials that exhibit a low work function and therefore easily emit electrons into vacuum form the basis of electronic devices used in applications ranging from satellite communications to thermionic energy conversion. W-Ba-O is the canonical materials system that functions as the thermionic electron emitter used commercially in a range of high power electron devices. However, the work functions, surface stability, and kinetic characteristics of a polycrystalline W emitter surface are still not well understood or characterized. In this study, we examined the work function and surface stability of the eight lowest index surfaces of the W-Ba-O system using Density Functional Theory methods. We found that under the typical thermionic cathode operating conditions of high temperature and low oxygen partial pressure, the most stable surface adsorbates are Ba-O species with compositions in the range of $Ba_{0.125}O$ to $Ba_{0.25}O$ per surface W atom, with O passivating all dangling W bonds and Ba creating work function-lowering surface dipoles. Wulff construction analysis reveals that the presence of O and Ba significantly alters the surface energetics and changes the proportions of surface facets present under equilibrium conditions. Analysis of previously published data on W sintering kinetics suggests that fine W particles in the size range of 100-500 nm may be at or near equilibrium during cathode synthesis, and thus may exhibit surface orientation fractions well-described by the calculated Wulff construction.

**Table of contents figure**

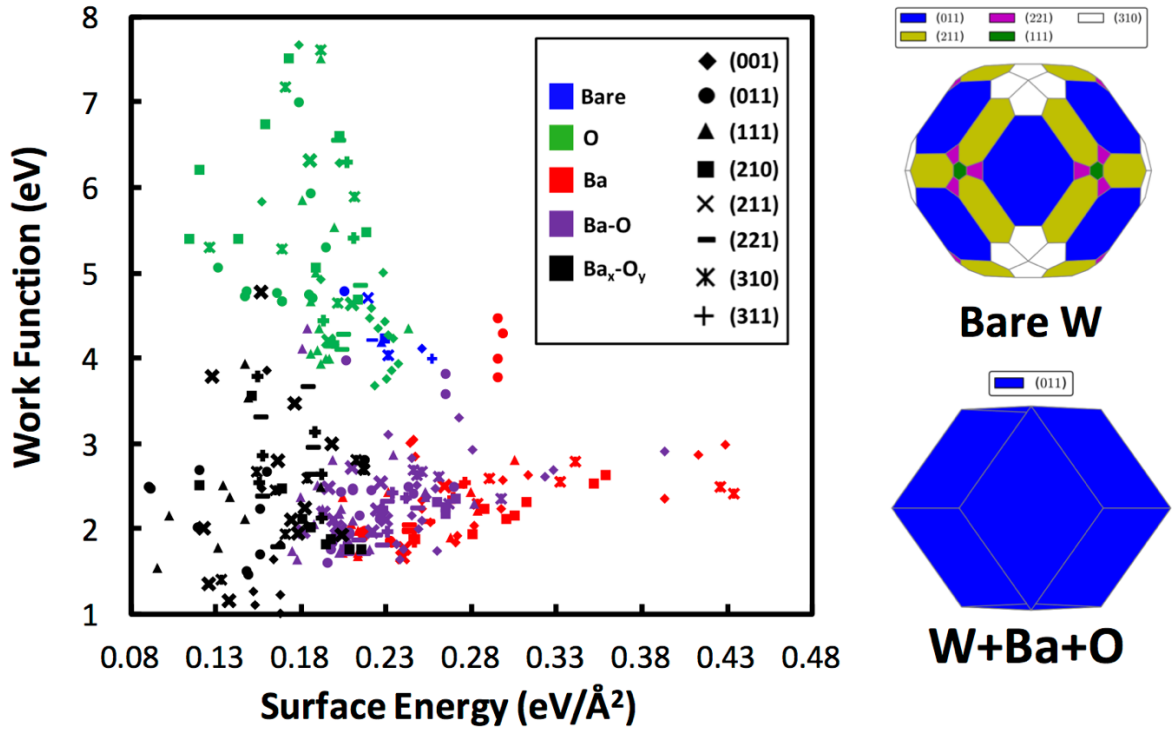

**Main**

Electron emission cathodes are found in high power, high frequency vacuum electronic devices (VEDs) such as traveling wave tubes, magnetrons and klystrons.[1] These high power VEDs are used in an array of applications, such as military and civilian infrastructure and communications, industrial food preparation, medical imaging, scientific research, and satellites.[2,3] All of these applications require low work function electron cathodes that provide ample electron emission to generate the electron beam necessary for the function of the VEDs. Recently, there has also been considerable interest in using electron emission cathodes in thermionic energy conversion devices.[4,5] In thermionic energy converters, the excess energy of an electron emitted into vacuum and re-absorbed by a material of lower work function results in a voltage difference capable of doing useful work; the excess energy of the hot electrons can also be coupled to a heat engine which can generate steam to power a turbine. In particular, the creation of hot electrons using solar energy could result in high efficiency, energy-generating VEDs that use thermionic electron cathodes.[4-7]

Researchers have published numerous studies detailing novel classes of materials spanning a large composition space that all show promise as low work function electron emitters in future VEDs. For example, adding $Sc_2O_3$ to traditional W-based cathodes (creating so-called scandate cathodes) lowers the work function, making scandate cathodes promising candidates for commercial high-power microwave VEDs.[8-14] Pure oxides have also been explored, e.g., perovskite oxides have been experimentally shown to function as a work function-lowering coating for field emitters,[15-17] and the perovskite work function physics and novel low work function surfaces have been investigated computationally.[18-20] Dichalcogenide materials have been a material class of interest for thermionic energy converters, with promising materials possessing predicted work functions less than 1 eV.[21] As a last example, work function engineering of two-dimensional materials has also garnered interest, with an investigation demonstrating graphene can exhibit work functions as low as 1 eV[22] and alloyed MXenes (two dimensional carbides and nitrides) have predicted work functions of about 1.5 eV.[23] As is clear from the examples provided here, materials for use as electron emitters include a variety of material classes. The successful engineering of these materials as electron emitters will require understanding the complex interplay of the composition, chemistry, structure, and resultant properties in a full device environment. To guide the development of next generation materials it is particularly important to better understand the systems that are already in use, like W-Ba-O.

W-Ba-O is a canonical thermionic cathode materials system used in commercial high power VEDs. Bare W has a high average work function on the order of 4.6 eV for a polycrystalline sample.[24] To be useful for device applications, the work function of W must be lowered, and a work function of about 2 eV can be achieved via the adsorption of Ba-O species on the emitting surfaces.[25-27] These Ba-O adsorbates produce electropositive dipoles which electrostatically reduce the work function directly at the emitting surface.[28,29] Numerous emission cathode materials systems have been shown or predicted to reduce the work function via electrostatic dipole creation at the surface. Examples include W, Mo, Ta, Re and Ni with adsorbed Cs,[30,31] (additional examples of work function lowering of other metals with adsorbed alkali metals are also summarized in Ref. 31), graphite with adsorbed Cs-I,[32,33] diamond with adsorbed H, Cs-O, and transition metals such as Cu, Ni, Ti and V,[34-38] GaAs and Ge with adsorbed Cs,[39,40] and W and $Sc_2O_3$ with adsorbed Ba-O.[12,13,41-43]

The emitting properties of W have been studied previously with Density Functional Theory (DFT)-based approaches, including the stability and work function of Ba, Sc and O on the (001) surface,[41] the expected crystal structures and electronic properties of Os-doped W,[44] and the stability and emission characteristics of W coated with various oxide films and adsorbed alkali metals.[45] However, an in-depth investigation examining the stability and work function of numerous crystal faces of W with adsorbed Ba-O is still missing. This study addresses that gap and provides in-depth computational examination of the W-Ba-O system. Our goals include understanding the stable W and W-Ba-O surfaces, including the stable Ba and O structures on W surfaces, and how these structures impact surface stability and work functions under physically relevant temperature and oxygen partial pressure conditions. An in-depth analysis of W-Ba-O can enable improved experimental cathode characterization and design, provide a framework to guide future examinations of more complex electron emitter materials systems, and generally enhance the basic understanding of electron emission from the W-Ba-O system, which forms the basis of most commercial thermionic electron emitters used in high power VEDs.

All calculations in this study were performed using Density Functional Theory (DFT) as implemented in the Vienna Ab-Initio Simulation Package (VASP).[46] The generalized gradient approximation (GGA) exchange and correlation functional with Perdew, Burke and Ernzerhof (PBE)-type pseudopotentials and the projector augmented wave (PAW) method were used for the W, Ba and O atoms.[47,48] The planewave energy cutoff was set to be 500 eV for all calculations. Wulff construction analysis was used to determine the predicted equilibrium shape of W under different thermodynamic environments.[49] All surface slabs and Wulff constructions were generated using the tools contained in the Pymatgen code package.[50] The starting structure used as input for making surface slabs with Pymatgen was the fully relaxed 2-atom conventional BCC W unit cell (2 atoms, space group $Im\bar{3}m$, $a=b=c=3.1847$ Å). The work function and surface energies for each surface configuration were calculated using methods well-documented in previous studies.[13,18,41,51] The reference states of O and Ba used for surface energy calculations were taken as the $O_2$ gas energy from the Materials Project[52] and rocksalt BaO, respectively, and our designation of "thermionic cathode operating conditions" corresponds to temperature (T) and oxygen partial pressure ($p(O_2)$) values of 1200 K and $10^{-8}$ Torr, respectively. Following previous studies, we apply a shift for the O chemical potential to account for the vibrational energy differences between O in the gas and surface adsorbed phase (this shift approximately cancels for the Ba chemical potential).[14,41] This was done by using an Einstein model with Einstein temperatures of 135, 130 and 1245 K for vibrations in the $x$, $y$ and $z$ directions (where $z$ is normal to the surface), respectively, obtained by simulating a single O atom bound to the (001) W surface using the finite-differences method. We examined adsorption of Ba atoms, O atoms, Ba

+ O in the ratio of Ba/O = 1 and Ba + O in the ratio of Ba/O < 1 on the (001), (011), (111), (210), (211), (221), (310) and (311) surfaces. These eight lowest index surfaces were examined to simultaneously explore a larger set of surface orientations than is typically done for DFT investigations of surface adsorption, and to keep the number and time of calculations tractable. In addition, there is currently no evidence that higher index surfaces than those explored here are stable under thermionic cathode operating conditions with adsorbed species present. We identified the stable adsorption sites of the Ba and O atoms for each surface by considering sites with the atoms directly on top of the surface W atoms, or midway between W nearest neighbors. For adsorption of Ba + O, we always placed the O atoms between the W and Ba, as this general configuration has been shown to yield more stable configurations than O on top of Ba.[41,42] For the specific adsorption case of Ba+O, different ratios of Ba/O were simulated (in increments of 1 Ba per 8 surface O) until the stable Ba/O ratio was found. A total of 295 surfaces were modeled in this work. The relaxed coordinates of all surface structures, as well as the surface slab sizes, adsorption compositions, calculated total energies, work functions, surface energies and surface electrostatic potential plots can be found in the **Supplementary Material**. To ensure convergence of the work function and surface energy, we simulated each surface slab with approximately 20 Å of vacuum space and relaxed the top and bottom first 4-5 Å of surface atoms, with the remaining atoms in the slab fixed to their bulk coordinates. We note here that the relaxation of the first 4-5 Å of surface atoms yielded well-converged work functions and surface energies for a test set of the (001), (011) and (111) surfaces, and therefore this criterion was adopted for all surfaces in this study. We modeled the *k*-point meshes with the Monkhorst-Pack scheme.[53] For each surface orientation, the *k*-point mesh was chosen following two general strategies: (1) the *k*-point values for each supercell axis were scaled inversely with the length of that axis to maintain a similar *k*-point sampling in reciprocal space and (2) the *k*-point values used for surface orientations other than (001) were scaled based on the *k*-points used for the (001) surface, again to maintain a similar *k*-point sampling in reciprocal space. We explicitly tested the *k*-point convergence of work function and surface energy for the (001), (011), (111), (210), (211), (221), (310) and (311) surfaces, and found that *k*-point meshes of 8×8×1, 8×4×1, 4×4×1, 3×6×1, 3×6×1, 6×3×1, 3×9×1 and 3×6×1 for the above listed surface orientations, respectively, resulted in convergence of the work function and surface energy to within approximately 0.05 eV and 0.001 eV/Å$^2$, respectively. We note there have been previous studies which calculated the surface energies of a series of bare W surfaces.[54,55] In particular, the work of Tran, et al. used the same methods to calculate the surface energies of bare W surfaces as used in this work.[54] The average difference between our calculated surface energies and their reported values is 2.4 ± 1.1 %, where the 1.1% spread is the standard error in the mean. This level of error is reasonable given differences in calculation parameters such as layer thickness, k-point mesh, vacuum layer thickness, and number of layers relaxed.

It is worth noting that we have only considered adsorption of Ba and O species on atomically flat W surfaces. Real polycrystalline W samples contain regions of flat surfaces, but may also contain surface steps, ledges, or pits where the surface structure is locally different from the atomically flat surfaces considered here, thus potentially producing adsorption structures different from those considered in this work. There have been previous experimental studies which have examined the emission of constrained adsorption of Ba and O on single crystal W surfaces, and emission from polycrystalline W surfaces which may exhibit more complex adsorption structures.[25-27] In both cases, the experimental work functions are very similar (effective thermionic work functions of about 2 eV), and previous DFT studies of Ba and O adsorption on W have found good agreement with these experimental results.[41,42] Given this agreement, we believe our examination of the work function and stability of adsorbates on atomically flat W surfaces serves as a sufficiently accurate model representation of the surfaces present in a polycrystalline W cathode measured in experiment.

**Figure 1** contains a plot of calculated work function as a function of surface energy for every surface orientation and adsorbate termination examined in this work. The different symbol types represent different surface orientations and the different colors denote different surface adsorbed species. Here, we describe the chemical trends in work function and surface energy for each adsorbate type.

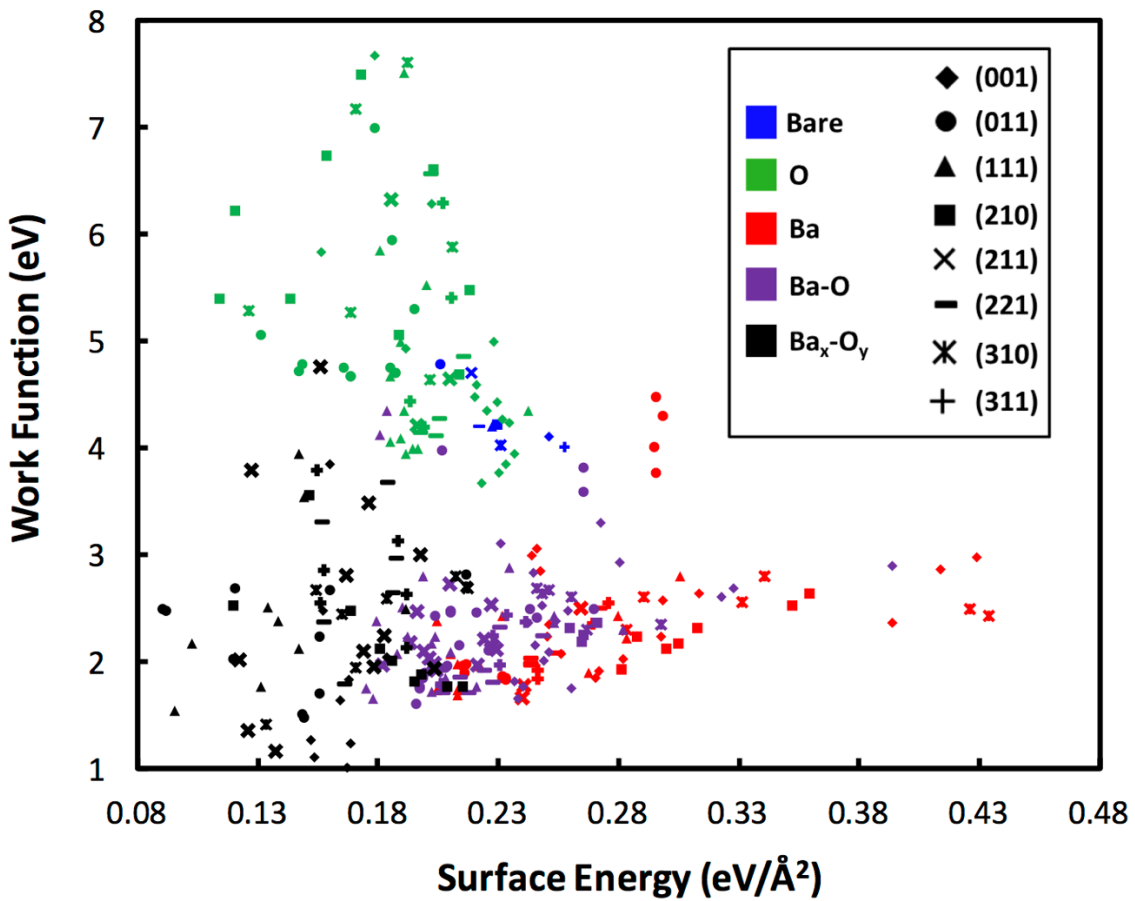

**Figure 1.** Work function versus surface energy (calculated under thermionic cathode operating conditions) for the (001), (011), (111), (210), (211), (221), (310) and (311) surfaces (symbol types) that are bare (blue points) or contain adsorbed O (green points), Ba (red points), Ba-O

where Ba/O =1 (purple points), $Ba_xO_y$ where x(Ba)/y(O) < 1 (black points). Data of all work functions and surface energies shown here can be found in the spreadsheet which is part of the **Supplementary Material**.

In **Figure 1**, the bare surfaces (blue data) tend to reside between work functions of 4-5 eV and middle to high values of surface energy. The bare work function values agree with known experimental values placing the polycrystalline W work function at about 4.6 eV. The relatively high bare surface energies compared to others on the plot indicate that W can lower its surface energy substantially by adsorbing O and partially oxidizing, or adsorbing Ba and O if a source of Ba is available.

The results for the O-adsorbed surfaces (green data) show that even under thermionic cathode operating conditions, W can lower its surface energy by forming a partial monolayer of O atoms, indicating that proper high vacuum control is crucial to avoid potential cathode poisoning. Due to their electronegative nature, adsorption of O atoms results in an increased work function.

The results for the Ba-adsorbed surfaces (red data) show that the majority of surfaces with adsorbed Ba are less stable than their bare variants. This indicates that metallic Ba will tend to desorb from the W surface to form BaO. In contrast to the adsorbed O surfaces, the electropositive nature of Ba substantially reduces the work function.

For the surfaces containing adsorbed Ba-O with a ratio of Ba/O = 1 (purple data), the range of work function values is similar to the case of adsorbed Ba, however the surface energy is decreased relative to pure Ba due to the presence of O. For the surfaces containing adsorbed Ba-O with a ratio of Ba/O < 1 (black data), the range of work functions is rather large, between 1-4 eV (higher work functions are the result of sparse coverage of Ba-O), and the surface energies are the lowest for all adsorbate species considered here. For the (001) surface, our most stable surface is the same $Ba_{0.25}O$ arrangement as reported previously by Vlahos, et al, who only investigated the (001) surface.[41] A key result of this analysis is that the O-rich Ba-O (Ba/O < 1) species result in the most stable surfaces not just for the (001) surface, but for *every* surface orientation investigated. This result is consistent with assigning formal valences of $Ba^{2+}$ and $O^{2-}$, assuming that some oxidation of the W takes place, and then balancing charge. For the (001), (011), (111), (211) and (310) surfaces, these stable Ba-O (Ba/O < 1) surfaces have low work functions of ≤ 2 eV. These findings indicate that both low and high index surfaces may exhibit a similar low work function and therefore could contribute appreciably to the measured electron emission.

Here, we discuss the structural and chemical characteristics which give rise to the most stable W-Ba-O structures for each surface orientation. **Figure 2** contains surface structures of the most stable simulated structures under thermionic cathode operating conditions for the (001) (**Figure 2A**), (011) (**Figure 2B**), (111) (**Figure 2C**), (210) (**Figure 2D**), (211) (**Figure 2E**), (221) (**Figure 2F**), (310) (**Figure 2G**), and (311) (**Figure 2H**) surfaces. For all surfaces, the most stable adsorbed species are Ba-O, where Ba/O < 1. The stable Ba-O compositions for each surface were found to be the following: (001): $Ba_{0.25}O$, (011): $Ba_{0.125}O$, (111): $Ba_{0.25}O$, (210): $Ba_{0.125}O$, (211): $Ba_{0.25}O$, (221): $Ba_{0.25}O$, (310): $Ba_{0.125}O$ and (311): $Ba_{0.125}O$, i.e., the stable Ba/O ratios for all surfaces examined were either 1/4 or 1/8. We note here that the true stable Ba/O ratio may deviate slightly from 1/4 or 1/8, as the range of the Ba/O ratio was not explored in finer increments than 1/8. In general, the stable structures all consist of the O atoms bonded to W, where the O atom position lies either directly on top of the W atoms (e.g., the (001) surface), on top of the W atoms but with a slightly bent bond (e.g., the (111) surface), or situated between nearest neighbor W atoms (e.g., the (011) surface). Additionally, the Ba atoms are bonded above the O to create an electropositive surface dipole, and reside between nearest neighbor O atoms and above subsurface W atoms. For every surface examined, the most stable Ba/O ratio resulted in the O atoms fully passivating the under-coordinated surface W atoms.

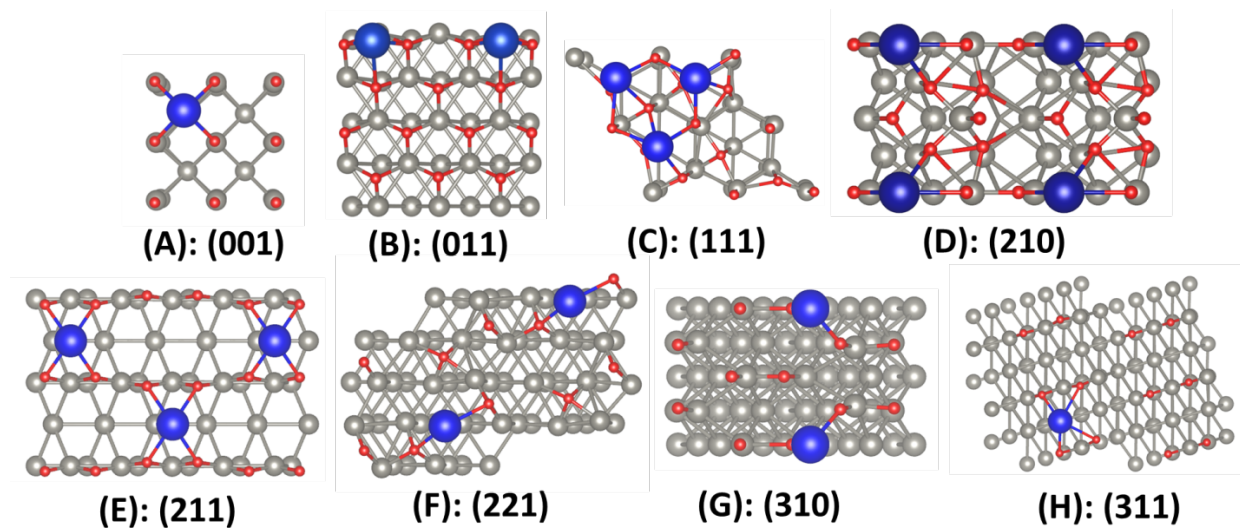

**Figure 2.** Surface structures depicting the most stable (lowest surface energy) adsorbed species for the (001) (A), (011) (B), (111) (C), (210) (D), (211) (E), (221) (F), (310) (G) and (311) (H) surface orientations under thermionic cathode operating conditions. The grey, red and blue spheres are W, O and Ba atoms, respectively. In all subfigures, the surface orientation direction is pointing out of the page. For all surfaces, the most stable adsorbed species is Ba-O where Ba/O < 1. More specifically, the stable surface compositions are: (001): $Ba_{0.25}O$, (011): $Ba_{0.125}O$, (111): $Ba_{0.25}O$, (210): $Ba_{0.125}O$, (211): $Ba_{0.25}O$, (221): $Ba_{0.25}O$, (310): $Ba_{0.125}O$, and (311): $Ba_{0.125}O$, respectively. All surface figures were generated using the VESTA code.[56]

To obtain additional insight into which surfaces may be present in an experimental W-Ba-O cathode and the expected proportion of each surface present at thermodynamic equilibrium, we used the calculated surface energies from **Figure 1** and calculated the Wulff construction under thermionic cathode operating conditions. **Figure 3** shows the Wulff constructions for four different physical situations: **Figure 3A** is the Wulff construction assuming an environment consisting of only pure W (i.e., perfect vacuum), **Figure 3B** assumes an environment of W + Ba, but devoid of O, **Figure 3C** assumes an environment consisting of W + O, and **Figure 3D** consists of W + Ba + O. **Table S1** of the **Supplementary Material** contains the computed area fractions and work functions of the various surface terminations present in each Wulff construction depicted in **Figure 3**. It is evident from **Figure 3** that the presence of O, Ba or Ba and O together profoundly alters the surface thermodynamics to affect which surface orientations are expected to be present in equilibrium, as well as the fraction of each orientation. This result suggests that the presence of different metal species on the cathode surface (for example, Ba or Sc in a scandate cathode), may impact the proportion of different surface orientations present in the cathode. The fraction of each orientation present will be affected by these surface species if the dynamics of W surface diffusion is sufficiently fast to allow at least nanoscale re-faceting of the W when exposed to the additional surface species, either during synthesis or operation, thus impacting the proportion of different surfaces present and the overall work function of the emitting surface. Here, we briefly remark on the expected work function values resulting from each equilibrium Wulff construction depicted in **Figure 3**. As multiple surface orientations may be present, the effective work function of a polycrystalline W cathode will be an average of the work function of each surface termination. For each Wulff construction, we have calculated the arithmetic-averaged work function and the exponential-averaged work function. The arithmetic-averaged work function was calculated by averaging the work functions of all the orientations present weighted by their area fraction on the Wulff construction. The exponential averaged work function was calculated by averaging the work functions of all the orientations present weighted by their thermionic current density at $T$=1200 K as calculated using the Richardson-Dushman equation, thereby obtaining an effective high temperature work function. These averaged work functions are provided in **Table S2** of the **Supplementary Material**.

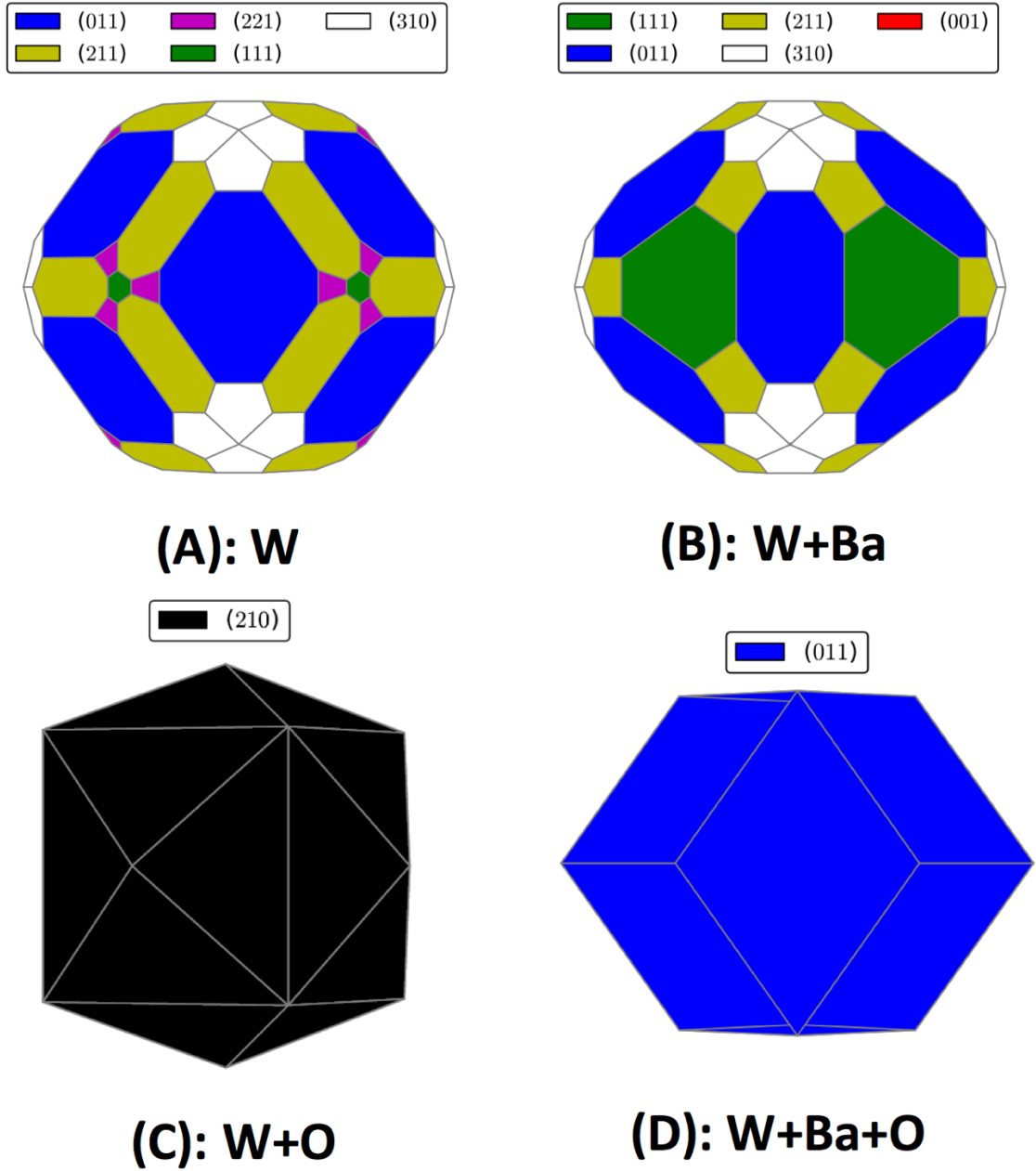

**Figure 3.** Calculated Wulff constructions indicating the prevalence of most stable surfaces under thermionic cathode operating conditions for the case of (A): pure W, (B): W with Ba, (C): W with O, and (D): W with Ba and O present.

To the extent that W-based cathode materials are at equilibrium, the above Wulff constructions should be consistent with observed surfaces. While there have been very few characterizations of cathode particle surface faceting of which we are aware, a few examples are available in the literature. The work of Wang, et al. produced scandate ($Sc_2O_3$-containing) W cathodes using a liquid-solid doping technique. With this technique, micron-sized tungsten oxide particles were suspended in solution with dissolved Sc (from $Sc_2O_3$), and subsequently annealed in a reducing atmosphere to evaporate the solvent and reduce tungsten oxide to metallic W.[57,58] From Figure 1 of their work (see Ref. 57), there is evidence of W particle faceting on a sub-micron scale. Similarly, the work of Vancil, et al. fabricated scandate cathodes using the same procedure as detailed by Wang, et al. An SEM micrograph from Vancil, et al. (see Figure 4 of Ref. 59) of the un-sintered powder clearly shows faceted W particles whose shapes contain the same surface terminations in qualitatively the same fraction as that for the W+Ba+O shape in **Figure 3D**.[59] We note here that the samples from Vancil, et al. contained W+Sc+O, and we have no Sc but do have Ba in our calculations. In addition, the lack of complete Sc-O coverage on the W particles observed by Vancil, et al. is not the same as the Ba-O equilibrium considered here, though the lack of Sc is a necessary but not sufficient condition to having equilibrium established by Ba-O. Therefore, our present comparison of W+Ba+O with the W+Sc+O system studied by Vancil, et al. is highly approximate and of only qualitative value, but is also the best comparison that can be made at present as no equivalent study comprising W+Ba+O has been performed. We speculate that the similarities between the W crystal shapes observed by Vancil, et al. and our calculated Wulff construction are due to Sc and Ba acting as reducing agents in similar ways, at least with regard to the surface energetics. Interestingly, there are also W particles visible in the SEM images in the work of Vancil, et al. and Wang, et al. that appear to be many small crystallites that have partially sintered together from the process of solvent removal during synthesis. The length scale of these partially sintered crystals is about 100-500 nm. Our observed qualitative agreement between experimental and predicted crystal shapes suggests that the kinetics of W diffusion among these 100-500 nm sized particles is sufficiently fast for the system to be at or near equilibrium during synthesis.

To further assess the expected length and time scales of W surface kinetics during sintering, we calculated diffusion lengths as a function of time for typical sintering temperatures[60-62] between 1500-1700 °C (see **Figure S1** of the **Supplementary Material**) using a basic Arrhenius relationship and activation barriers for W diffusion during sintering obtained from the literature (see **Supplementary Material** for more details).[63-65] Based on the work of Kothari, et al.,[63] Vasilos, et al.,[65] and German, et al.,[64] who all examined the diffusion rates of W during sintering, the average activation barrier to diffuse W along a polycrystalline surface is about 4.57 eV. Thus, at a typical sintering temperature of 1500 °C (1700 °C), a typical diffusion length may be on the order of 25 (120), 36 (160) and 50 (230) nm for typical times of 30, 60 and 120 minutes of sintering, respectively. For 1700 °C, these length and time scales are qualitatively consistent with the previous observation that W particles in the submicron size range of 100-500 nm may be at or near equilibrium. However, we note here that many emitter cathodes use W powders with particle sizes of 1 μm or larger, and it would take approximately two days of sintering for W to move a distance of about 1 μm at 1700 °C, which is much longer than typical sintering times (to equilibrate with a W diffusion length of about 1 μm in a sintering time of one hour would require sintering at 2050 °C, a much higher temperature than what is typically used to sinter tungsten). Thus, larger grain sizes may result in sintered grains whose fractions of surface facets deviate from those predicted by the Wulff construction as a result of kinetic limitations of W surface diffusion during the sintering process. In addition, other non-equilibrium processing steps, such as machining, etching or cleaning, may result in the presence of surface terminations that are not expected based on thermodynamic predictions alone.

Therefore, the processing steps of preparing W powders used to construct thermionic dispenser cathodes may significantly affect which surface orientations are present, directly impacting the value of the overall measured work function and performance of the cathode.

In summary, we used DFT methods to analyze the work function and surface stability of the eight lowest index terminations of W in the presence of Ba and O adsorbates. We found that Ba-O adsorbates where Ba/O < 1 are the stable adsorbates for all surfaces under thermionic cathode operating conditions. Further, we found that thermodynamics favors passivating each dangling W bond with a single O atom, followed by adsorption of Ba to produce compositions of approximately $Ba_{0.125}O$ or $Ba_{0.25}O$. For numerous surfaces, these adsorbates produce work functions around 2 eV or below, in close agreement with experimental measurements of the effective work function. Wulff construction analysis showed that the presence of Ba and O significantly alter the proportion of surface terminations present at equilibrium, with a system comprising W, Ba and O exhibiting mainly (011) and (111) terminations under cathode operating conditions. Finally, we used previously published data of W sintering kinetics to show that the precise sintering temperature, time, and W particle size may play a significant role in setting the fraction of surfaces present in a real cathode. The results and methods employed in this work may directly influence experimental and computational investigations of other thermionic cathode systems such as scandate, Os-Ru, or oxide cathodes, and offer basic investigative principles useful for Cs-coated metal or semiconductor cathodes used extensively in thermionic conversion devices.

**Supplementary Material:** See supplementary material for calculated data of work function and surface energy for all surfaces examined in this study, essential calculation input files and final structures, and an analysis of W diffusion during sintering.

**Acknowledgements:** This work was supported by funding from the Defense Advanced Research Projects Agency (DARPA). This research used computing resources of the National Energy Research Scientific Computing Center (NERSC), which is supported by the U.S. Department of Energy Office of Science. This research was also performed using the compute resources and assistance of the UW-Madison Center for High Throughput Computing (CHTC) in the Department of Computer Sciences.